# Controllable chaotic dynamics in a nonlinear fiber ring resonators with balanced gain and loss

Jyoti Prasad Deka . Samit Kumar Gupta . Amarendra K. Sarma*

**Abstract** We show the possibility of controlling the dynamical behavior of a single fiber ring (SFR) resonator system with the fiber being an amplified (gain) channel and the ring being attenuated (loss) nonlinear dielectric medium. Our model is based on the simple alterations in the parity time symmetric synthetic coupler structures proposed recently [A. Regensburger et al., Nature **488**, 167 (2012)]. The system has been modeled using the transfer matrix formalism. We find that this results in a dynamically controllable algorithm for the chaotic dynamics inherent in the system. We have also shown the dependence of the period doubling point on the input amplitude, emphasizing on the dynamical aspects. Moreover, the fact that the resonator essentially plays the role of a damped harmonic oscillator has been elucidated with the non-zero intensity inside the resonator due to constant influx of input light.

**Keywords** Chaos. Ikeda. Fiber coupler. Resonators

J. P. Deka . S. K. Gupta . A. K. Sarma
Department of Physics, Indian Institute of Technology Guwahati, Guwahati-781 039, Assam, India
*aksarma@iitg.ernet.in

**1 Introduction**

Chaotic systems exhibit sensitivity to the initial conditions. In such systems any uncertainty, no matter how small, in the beginning will produce rapidly escalating and compounding errors in the prediction of the system's future behavior. It implies that two trajectories emerging out from two distinct but nearby initial conditions diverge exponentially from each other with the passage of time. The idea of chaos is studied quite extensively in natural sciences, especially in chemistry [1-3], electronics [4-5], fluid dynamics [6] and nonlinear optics [7-11].

In this work, we study a single fiber resonator (SFR) system with balanced gain and loss modeled by the so-called Ikeda map [12]. It should be noted that, a discrete time dynamical system known as the Ikeda map, proposed by K. Ikeda, model optical dynamics in nonlinear optical resonators [12, 13]. This gives rise to chaotic dynamics in some specific parametric regime.

In this context, dynamical pulse shaping and period-doubling bifurcations have been studied in nonlinear fiber ring resonator systems both theoretically [14-16] and experimentally [17-20]. Here, we study the effects of the balanced gain and loss profile upon the evolution dynamics of the input optical pulse over a sufficiently large number of round-trips in the ring cavity. It is found that input light above a certain threshold power can give rise to chaos due to the interference between the input field and the cavity field. The article is organized as follows. In Sec. 2, we present a discussion on the formalism used and the theoretical model. Sec.3 contains the results and discussions followed by conclusions in Sec. 4.

**2 The Model**

In an optical fiber coupler, the coupled mode equations governing the dynamics of the field amplitudes are given by:

$$i\frac{da_1}{dz} = \beta_1 a_1 + C a_2$$

$$i\frac{da_2}{dz} = \beta_2 a_2 + C a_1 \quad (1)$$

Here, $z$ is the propagation distance, $a_1$ and $a_2$ are field amplitudes in fiber 1 and fiber 2 respectively. $\beta_1$ and $\beta_2$ are the propagation constants of the two waveguides and $C$ is the coupling coefficient. For identical fibers, $\beta_1 = \beta_2$ and analytical solution of these equations yield the transfer matrix as follows:

$$\begin{pmatrix} a_1(z) \\ a_2(z) \end{pmatrix} = \begin{pmatrix} \cos(Cz) & i\sin(Cz) \\ i\sin(Cz) & \cos(Cz) \end{pmatrix} \begin{pmatrix} a_1(0) \\ a_2(0) \end{pmatrix} \quad (2)$$

The transfer matrix can be rewritten, defining the intensity splitting ratio, $k \equiv (\sin(CL))^2$ where $L$ is the coupling length, as follows [21]:

$$A = \begin{pmatrix} \sqrt{1-k} & i\sqrt{k} \\ i\sqrt{k} & \sqrt{1-k} \end{pmatrix} \quad (3)$$

For a 50:50 directional coupler, $k = \frac{1}{2}$. Hence, the transfer matrix for 50:50 directional coupler is given by:

$$A = \frac{1}{\sqrt{2}}\begin{pmatrix} 1 & i \\ i & 1 \end{pmatrix} \qquad (4)$$

The fiber coupler considered in this work is inspired by the parity-time (PT) synthetic coupler introduced in Ref. [22, 23]. The coupler comprises a passive coupling region and two channels of waveguides, one of which provides amplification and the other provides an equal amount of attenuation. It is worthwhile to note that optical fibers can be synthesized to amplify as well as attenuate optical power by suitably tailoring their refractive index profile. Suppose the refractive index profile of an optical fiber is $n = n_r + in_I$. If $n_I < 0$, the light input will undergo amplification, whereas if $n_I > 0$, it will suffer attenuation. The two channels are connected to the passive coupling region as shown in Fig.1. This configuration served as the building block of *PT* symmetric optical mesh lattices in the aforementioned works. In this work, we consider the same *PT synthetic coupler* but with a slight modification (as shown in Fig.1).

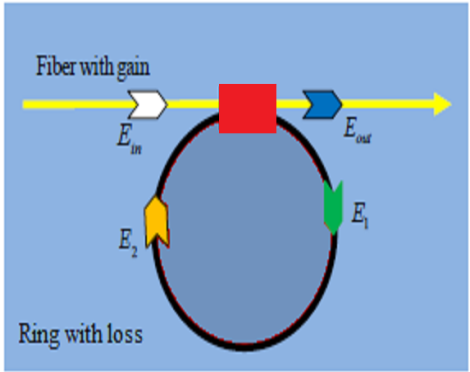

**Fig. 1** (Color online) Schematic diagram of the SFR resonator structure. The yellow-color region represents the fiber with gain, the black-color ring represents the resonator with loss and the red-color rectangular block is the passive coupling region. $E_{in}$ is the input field, $E_{out}$ the output field and $E_1$ is the part of input field transferred via the ring resonator back to the coupling region as $E_2$.

The system has been modified in such a way that the output port from the loss channel has been fed back into its input. And we consider the loss channel to be a nonlinear dielectric medium. This provides us with a loss resonator channel and we can write the transfer matrix as follows:

$$M = \begin{pmatrix} e^{\frac{\gamma}{2}} & 0 \\ 0 & e^{-\frac{\gamma}{2}} \end{pmatrix} \frac{1}{\sqrt{2}}\begin{pmatrix} 1 & i \\ i & 1 \end{pmatrix}\begin{pmatrix} e^{\frac{\gamma}{2}} & 0 \\ 0 & e^{-\frac{\gamma}{2}} \end{pmatrix} \qquad (5)$$

Here, $'\gamma'$ is defined as the loss parameter. And it must be noted here that there is no evanescent coupling between the optical fibers. It can be seen from Eq.5 that the matrix in the middle is that of 50:50 directional coupler and the matrix on the left and right are that of two optical fibers, one of which amplifies light amplitude by a factor of $e^{\frac{\gamma}{2}}$ and the other attenuates the same by a factor of $e^{-\frac{\gamma}{2}}$. In a nutshell, we can say that our system couples an amplified and an attenuated light input via a 50:50 directional coupler and redirects one output back into directional coupler after it has traversed a lossy nonlinear resonating channel and the other output is sent to a detector to study the evolution dynamics in the system.

Using the transfer matrix $M = \frac{1}{\sqrt{2}}\begin{pmatrix} e^{\gamma} & i \\ i & e^{-\gamma} \end{pmatrix}$, we can write the output amplitudes in terms of the input amplitudes as follows:

$$\begin{pmatrix} E_{out} \\ E_1 \end{pmatrix} = M \begin{pmatrix} E_{in} \\ E_2 \end{pmatrix} \qquad (6)$$

$E_{in}$ and $E_2$ are the input field amplitudes to the 50:50 directional coupler and $E_{out}$ and $E_1$ are the output field amplitudes to the same. The field $E_1$ suffers a nonlinear phase shift $\emptyset_{NL}$, when it travels one complete loop across the ring. So, we can write: $E_2 = E_1 e^{i\emptyset}$ where $\emptyset = \emptyset_L + \emptyset_{NL}$ is the total phase shift suffered by the field. $\emptyset_L$ is the linear phase shift. The round-trip time for one complete loop of the electromagnetic wave around the resonator is given by: $t_R = nL/c$. $L$ is the length of the resonator, $'n'$ is the effective refractive index in the resonator. We are considering a Kerr medium in this model and the nonlinear phase shift, $\emptyset_{NL}$ that be imparted by such a medium is given by [24]:

$$\emptyset_{NL} = \frac{2Ln_2\pi}{A_{eff}\lambda_o}|E_k|^2 \qquad (7)$$

Here, $k = 1,2$ and $n_2$ is the nonlinear refractive index coefficient, $A_{eff}$ is the effective core area of the fiber and $\lambda_0$ is the wavelength of the propagating light in vacuum. Using Eq. (6) we can write the discretized field amplitude evolution equation in the resonator taking time steps equal to $t_R$ in the form of an iterative equation:

$$E_{j+1} = A + BE_j exp(i(\emptyset_{NL} + \emptyset_L)) \quad (8)$$

where $A = iE_{in}/\sqrt{2}$ and $B = e^{-\gamma}/\sqrt{2}$. This equation relates the resonator field amplitude at $(j+1)$th iteration with that of the $j$th iteration. Without any loss of generality, taking $\emptyset_L = 0$, we can rewrite Eq.8 as:

$$E_{j+1} = A + BE_j e^{i|E_j|^2} \quad (9)$$

In Eq. 9, we have considered the coefficient before $|E_j|^2$ as unity since this facilitates us to study the dynamical behavior of the system without a prior choice of the parameters of the ring resonator. From Eq. 6, we can obtain:

$$E_1 = \frac{iE_{in}+e^{-\gamma}E_2}{\sqrt{2}} \quad (10)$$

Now we know that, $E_2 = E_1 e^{i\emptyset}$. We can use this to further simplify our expression for $E_1$ as follows:

$$E_1 = \frac{iE_{in}}{\sqrt{2} - e^{-\gamma}e^{i\emptyset}}$$

Hence, the field intensity in the resonator is given by

$$I_{resonator} = \frac{|E_{in}|^2}{e^{-2\gamma}+2-2\sqrt{2}e^{-\gamma}\cos(\emptyset)} \quad (11)$$

At resonance, $\emptyset = 2\pi m$, where $m$ is an integer. Using Eq. (7) and setting $g = \frac{n_2}{A_{eff}\lambda_0}$, the resonance condition is found to be: $m = Lg|E_2|^2$. Here $g$ is the nonlinear parameter.

## 3 Results and Discussions

The system displays rich behavior in the context of nonlinear dynamics. The number of round trips (or iterations) light takes in the resonator shows how the field intensity evolves in the resonator. This evolution is governed by Eq. (6). Now because the ring is an absorbing medium, the intensity gets attenuated in it. As the iteration progresses, the system slowly enters the steady state. Now depending on the value of $\gamma$, the ring can either be in a chaotic state or stable, which are primarily the fixed points of Eq. (6). Taking a fixed value of $\gamma$, we carried out 1000 iterations and discarded the first 900 as transients and plot the remaining 100 steady state iterations. Fig.2 depicts the output intensity as a function of the input amplitude for $\gamma = 1.0$.

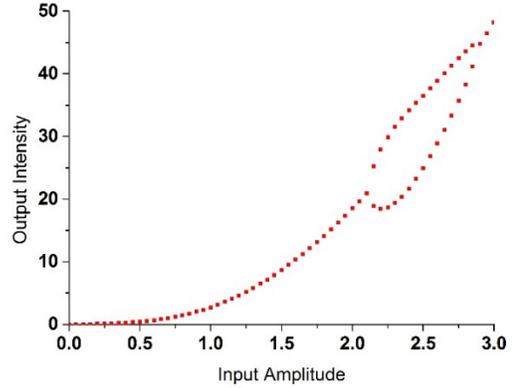

**Fig. 2** (Color online) Field Intensity in the Output Port vs. Input Field Amplitude for $\gamma = 1.0$.

It can be seen that the output intensity gets highly amplified as input amplitude increases. The instantaneous light intensity in the output port is always unique and well defined up to a certain range of the input amplitude. However, for a certain range of $E_{in}$, bifurcation behavior is observed. It is worth noting that if we increase $E_{in}$, the bifurcation behavior will cease to exist beyond a certain limit. On the other hand, if we proceed to increase $E_{in}$ beyond 3.0, as depicted in Fig. 3, we encounter a chaotic region beyond a certain point because $\gamma = 1.0$ is not sufficient to counteract the chaotic behavior of the system. The reason for this can be attributed to the fact that the ring resonator in our configuration is an intensity absorber.

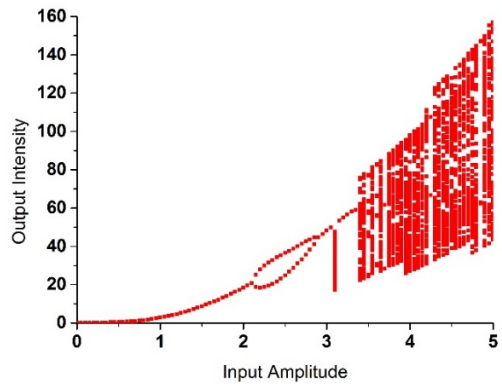

**Fig. 3** (Color online) Output intensity versus the input amplitude plot showing the chaotic behavior of the system for $E_{in}$ beyond 3.0.

In Fig.3, we can see that up to certain input amplitude, the resonator can attenuate the influx of energy and prevent the system from going to a chaotic state. But beyond a certain value, it fails in this aspect and we observe chaotic transmission of field intensity in the output port. The emergence of chaos is related to the input power and it may be understood as follows. The ring is being constantly pumped with energy. But because it is a lossy medium, it can attenuate the influx of energy. On top of that, we can see from Eq. 9 that the right hand side of the equation consists of two parts. The first term containing $'A'$ tells us about the influx of energy, whereas the second term tells us that the field amplitude has suffered attenuation as well as nonlinear phase shift when it traverses the whole length of the resonator. It is then superimposed with the incoming energy and fed back into the resonator. Now if the input power exceeds a certain limit, the lossy ring fails to balance the incoming energy with that of the energy flowing in the ring and this leads to the emergence of chaos. The role of the loss parameter $\gamma$ can be illustrated by plotting a bifurcation diagram of intensity in the resonator against the loss parameter $\gamma$, as shown in Fig. 4. As $\gamma$ decreases, the resonator field intensity has two distinct values, which goes to four on further decrease in $\gamma$ and eventually we enter into a region of chaotic behavior. It is clearly evident that the period doubling point lies close to $\gamma = 1.2$. $\gamma$ plays a crucial role in the transition of the system to a chaotic state. We can claim that $\gamma$ plays the role of a chaos control parameter. A plot of the largest Lyapunov exponent against the loss parameter γ, shown in Fig.5, validates our above-mentioned claim.

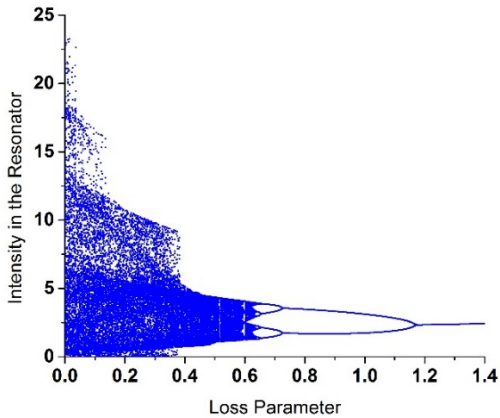

**Fig. 4** (Color online) Bifurcation diagram of the Resonator Field Intensity vs. $\gamma$ for $E_{in} = 2.5$.

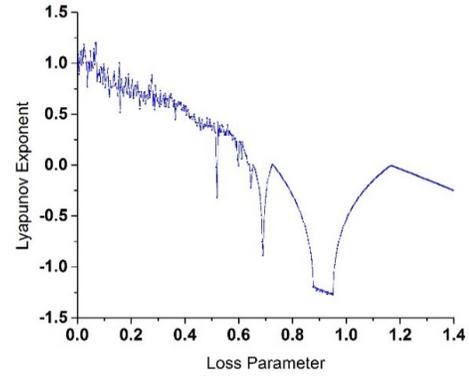

**Fig. 5** (Color online) Largest Lyapunov exponent vs. $\gamma$ for $E_{in} = 2.5$.

For $\gamma$ above 1.2, the system has negative Largest Lyapunov exponent. This implies that our chosen input field amplitude $E_{in} = 2.5$ corresponds to a stable state of the system, as is evident from the bifurcation diagram. In the region of $\gamma$ from 0 to 0.6, the Lyapunov exponents are positive indicating that the system is in chaotic state.

Now, a numerical algorithm can be designed which computes the value of $\gamma$ at which period doubling takes place for different values of $E_{in}$. This presents us with a set of $\gamma$ corresponding to the input amplitude $E_{in}$, as depicted in Fig. 6.

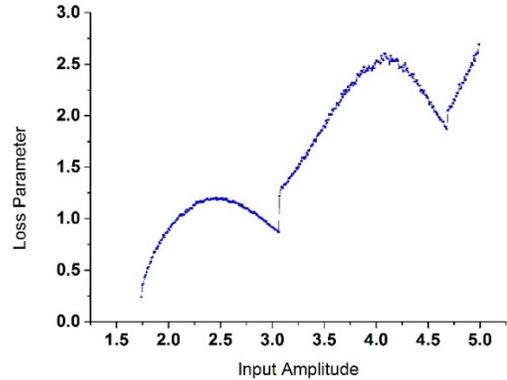

**Fig. 6** (Color online) Period doubling point vs. $E_{in}$.

The period doubling point shifts its value in accordance with $E_{in}$. Fig. 6 gives us an idea about how the system should be designed so as to observe a single intensity in the output. A single stable intensity in the resonator will ensure the same in the output. The period doubling point as calculated numerically for each value of $E_{in}$ corresponds to a maximal limit of $\gamma$, beyond which fluctuation in the resonator field intensity has been observed to be on the order of

$10^{-5}$. Some of the values of $E_{in}$ and $\gamma$ used to plot Fig. 6 are (2.0, 0.90), (2.5, 1.19), (3.0, 0.92), (3.5, 1.90) and (4.0, 2.53). Typically, for a Gaussian modal distribution, if the nonlinear refractive index $n_2$ is chosen to be on the order of $10^{-20}\ m^2W^{-1}$, $g$ takes values in the range of $10^{-3} - 10^{-2}\ W^{-1}/m$. Moreover, the nonlinear parameter $g$ can be increased by decreasing $A_{eff}$. Taking these parameter values under consideration, the resonance spectrum can be estimated using Eq. (7). Numerically, $g = 0.92$ has been found to restore the resonator intensity dynamics into a steady state when $E_{in} = 3.0$. The resonance spectrum for this particular case has been shown in Fig.7, which depicts the resonance profile with wavelength range chosen in the visible spectrum. The resonator field intensity peaks around 550-600 nm indicating that this configuration, if properly engineered, can be utilized for practical purposes.

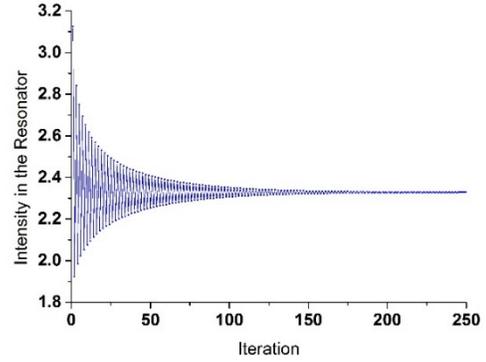

Fig. 8 (Color online) Intensity Evolution in the resonator vs. iteration for $E_{in} = 2.5$ and $\gamma = 1.19$.

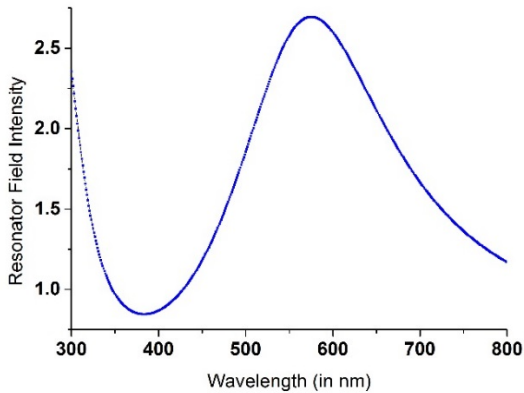

Fig. 7 (Color online) Figure showing the resonance profile for $E_{in} = 3.0$ and $\gamma = 0.92$. The parameters chosen are: $n_2 = 2.6 * 10^{-20}\ m^2W^{-1}, A_{eff} = 10^{-14}m^2\ and\ L = 0.5m$.

In Fig. 8 the intensity in the resonator has been plotted against the number of iterations of the optical fields. It can be seen that the resonator behaves like a damped harmonic oscillator, but the intensity does not decay down to zero owing to the fact that there is a constant influx of light, which is being fed to the resonator. The loss ring plays the role of a dampening medium and the input amplitude acts as the driving force in the system. Moreover, we have considered the ring to be a nonlinear dielectric medium. These elements are also a part of the pulse driven nonlinear oscillator equation [25], from which the Ikeda map is derivable and this justifies the behavior shown in Fig. 8. Hence, we can infer from the above plots that $\gamma$ plays the role of a chaos control parameter in our configuration. This work so far, on firm grounds, has validated the fact that the idea of a *PT* synthetic coupler can be converted into a simple fiber ring resonator with a control parameter that is solely dependent on the fiber characteristics. In addition to this, the fluctuations in the field intensity at the output port could be controlled with a judicious choice of the value of $\gamma$. The idea proposed in this work is general in nature. Hence, it could be exploited in any system having balanced loss and gain. As a specific example, it may be possible consider two inductively coupled LRC circuits, one with gain and the other with loss. It may be possible to control and manipulate chaos in such systems with judicious choice of parameters as described in our work.

## 4 Conclusions
In conclusion, we find that simple alterations in the parity time symmetric synthetic coupler structures could result in a dynamically controllable algorithm for the chaotic dynamics inherent in the system. We have also shown the dependence of the period doubling point upon the input amplitude, emphasizing on the dynamical aspects. Moreover, the fact that the resonator essentially plays the role of a damped and driven harmonic oscillator has been elucidated with the non-zero intensity inside the resonator due to constant influx of input light.

## Acknowledgments

J.P.D. and A.K.S. would like to acknowledge the financial support from DST, Government of India (Grant No. SB/FTP/PS-047/2013). S.K.G. thanks MHRD, Government of India for supporting through a fellowship.